\documentclass[reprint,groupedaddress,amsmath,amssymb,aps,prl]{revtex4-2}

\usepackage{graphicx}
\usepackage{bm}
\usepackage[normalem]{ulem} 

\usepackage{hyperref}

\hypersetup{
    breaklinks=true,
    colorlinks=true,
    allcolors=blue
}

\begin{document}
\title{Conditional Ergodicity and Universal Fluctuations in Weak Ergodicity Breaking}

\author{Dan Shafir}
\email{dansh5d@gmail.com}
\author{Stanislav Burov}
\email{stasbur@gmail.com}

\affiliation{Physics Department, Bar-Ilan University, Ramat Gan 5290002,
Israel}


\begin{abstract}
Time averages extracted from single-particle trajectories in complex media often vary strongly from one trajectory to another, even for long measurement times. Such persistent trajectory-to-trajectory scatter is commonly observed in anomalous diffusion and signals weak ergodicity breaking driven by scale-free trapping. Here we identify conditional ergodicity: conditioning on a natural internal clock restores self-averaging of time-averaged observables. Combining conditional ergodicity with the stochastic mapping between the internal clock and physical time implies a universal law: once rescaled by their mean, time-averaged transport coefficients in systems exhibiting weak ergodicity breaking follow the Mittag-Leffler distribution.
We demonstrate this universality across multiple models of disordered media displaying anomalous diffusion.
\end{abstract}

\maketitle

One of the fundamental concepts in statistical physics is the ergodic hypothesis, which states that for an ergodic system the long time average (TA) $\overline{\mathcal{O}(t)}$ of an observable $\mathcal{O}$ along a single trajectory equals its ensemble average (EA) $\langle \mathcal{O} \rangle$.
Namely, 
\begin{equation}\label{eq:ergodicity}
\overline{\mathcal O(t) }\equiv \frac{1}{t}\int_{0}^{t}\mathcal O(t')\,dt',
\qquad 
\overline{\mathcal O (t)}\xrightarrow[t\to\infty]{}\langle \mathcal O\rangle .
\end{equation}
The non-equivalence of $\overline {\mathcal{O}(t)}$ and $\langle {\cal O}\rangle$, known as ergodicity breaking, can arise when the dynamics effectively explores only part of the accessible state space over the measurement time. 
For example in the ferromagnetic phase of the Ising model the temporal evolution fails to sample the entire state space and depends on initial conditions.  
In contrast, weak ergodicity breaking (WEB), introduced by Bouchaud~\cite{bouchaud1992weak}, describes situations where the evolution is not confined to a specific region of state space, yet $\overline {\mathcal{O}(t)}$ remains trajectory dependent even as $t\to\infty$.
WEB often arises from scale-free trapping, where sojourn times exhibit a power-law tail with a divergent mean, preventing TAs from converging to a well-defined constant~\cite{bel2005weak,rebenshtok2007distribution}.
In this work, we investigate systems that exhibit WEB and demonstrate that identifying a natural internal clock reveals what we term conditional ergodicity: while TAs over a measurement window of duration $t$ vary across trajectories, conditioning on the internal clock collapses this trajectory-to-trajectory scatter.

\begin{figure*}[!ht]
\centering
\includegraphics[width=\textwidth]{"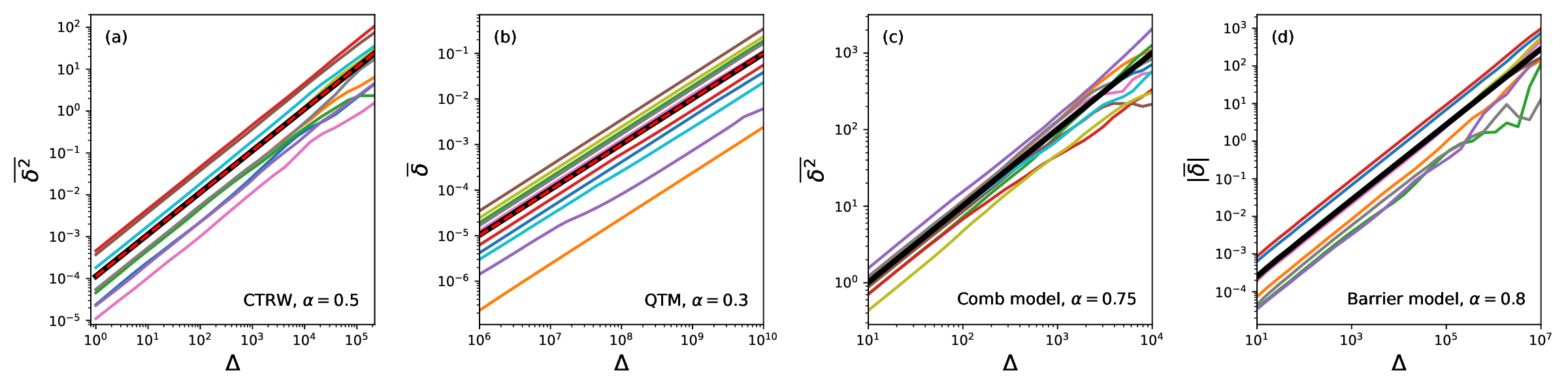"}
\vspace{-13pt}
\caption{ 
\textit{The scatter of TA observables across different models.}
The spread between trajectories indicates non-ergodic behavior for the models discussed: (a) TA-MSD in CTRW ($t=10^8$), (b) TA-displacement, $\overline{\delta(\Delta,t)} \equiv (t-\Delta)^{-1}\int_{0}^{t-\Delta} [x(t'+\Delta)-x(t')]\,dt'$, in transient QTM ($F = 0.1, \, t=10^{14}$), (c) TA-MSD in Comb model ($t=10^5$), (d) absolute TA displacement, $|\overline{\delta(\Delta,t)}|$, in Barrier model ($t=10^8$). For each model, 10 examples of the TA are given, color-coded.
The thick black line corresponds to the numerical average over 1000 simulated trajectories and the red dashed line in (a) and (b) are the known analytical solution - $\langle \overline{\delta^2} \rangle \sim \Delta / [A t^{1-\alpha} \Gamma(1 + \alpha)]$ for the symmetrical CTRW in $1d$~\cite{he2008random, lubelski2008nonergodicity} and $\langle \overline{\delta} \rangle \sim \Delta / [A \Lambda t^{1-\alpha} \Gamma(1 + \alpha)]$ (see Eq.~(S.11) in the SM~\cite{SM}) for the transient QTM in $2d$ \cite{shafir2022case}.
All models exhibit the same trajectory-to-trajectory scattering of the TA observable, as captured by Eq.~\eqref{eq:MittagLefflerDistribution}.
}
\label{Fig:1}
\end{figure*}

The WEB and the closely related phenomena of aging and non-self-averaging~\cite{bouchaud1992weak, monthus1996models, rinn2000multiple, rinn2001hopping, bertin2003subdiffusion, burov2007occupation} have evolved from a theoretical concept into an experimentally accessible feature of single-particle tracking.~\cite{scholz2016cycling, metzler2014anomalous, barkai2012strange, kompella2024determines, yu2018subdiffusion, anderson2019filament, sabri2020elucidating, wong2004anomalous, mckinley2009transient}. 
In such experiments, transport in disordered media is commonly quantified using TA observables, most notably the time-averaged mean-squared displacement (TA-MSD), 
\begin{equation} \label{eq:delta_definition}
\overline{\delta^2(\Delta, t)} \equiv \frac{1}{t-\Delta} \int_0^{t-\Delta}\left[x\left(t'+\Delta\right)-x\left(t'\right)\right]^2 \mathrm{~d} t'.
\end{equation}
$\overline{\delta^2(\Delta, t)}$ measures the squared displacement over a lag time $\Delta$ within a measurement window $t$, and leverages the whole trajectory to reduce noise.
A standard way to summarize each trajectory is via an apparent diffusion coefficient $\overline{D}$, extracted from the TA-MSD, $\overline{\delta^2(\Delta,t)} \simeq 2\,\overline{D}\,\Delta$ for small lag-time $\Delta$.
For ergodic dynamics, different trajectories yield similar $\overline{D}$ and $\overline{\delta^2(\Delta,t)}$ converges to the ensemble MSD at long $t$.
Under WEB, Eq.\eqref{eq:ergodicity} fails: $\overline{D}$ (and equivalently $\overline{\delta^2(\Delta,t)})$ remains broadly distributed across trajectories even for long measurement times [Fig.\ref{Fig:1}].
Such behavior has been reported in biological systems~\cite{luo2024quantum, scholz2016cycling, metzler2014anomalous, barkai2012strange, weigel2011ergodic, kompella2024determines, yu2018subdiffusion, anderson2019filament, sabri2020elucidating}, granular materials~\cite{mema2020fluidization, marty2005subdiffusion, bodrova2024diffusion, metzler2022modelling}, non-Newtonian fluids~\cite{wong2004anomalous} and other complex systems~\cite{mckinley2009transient, paoluzzi2024flocking}. 

WEB is often studied using the continuous-time random walk (CTRW)  model~\cite{scher1975anomalous, bouchaud1990anomalous}, where a scale-free power-law distribution of sojourn times with diverging mean leads to non-equivalence of the TA-MSD and ensemble  MSD~\cite{lubelski2008nonergodicity,he2008random,sokolov2008statistics}.  
In Ref.\cite{he2008random}, it was shown that, for CTRW, the diffusion coefficients extracted from single-trajectory TAs exhibit pronounced trajectory-to-trajectory scatter described by a Mittag–Leffler distribution. 
However, the widespread variability of transport coefficients observed in single-particle tracking experiments~\cite{luo2024quantum, scholz2016cycling, metzler2014anomalous, barkai2012strange, weigel2011ergodic, kompella2024determines, yu2018subdiffusion, anderson2019filament, sabri2020elucidating, mema2020fluidization, marty2005subdiffusion, bodrova2024diffusion, metzler2022modelling, wong2004anomalous, mckinley2009transient, paoluzzi2024flocking} indicates that  
the origins of this variability are generic and  
extends far beyond the simplistic assumptions of CTRW, which neglects quenched disorder, correlations, and geometric constraints. 
We therefore explore WEB in various models of disordered media: the quenched trap model, the comb model, and the random barrier model. 
By separating physical time from an internal clock via subordination~\cite{klafter2011first, weiss1994aspects} and using conditional ergodicity as a unifying framework, we obtain a universal distribution for the scatter of TAs. 
Despite the distinct transport properties of these models, the variability of TAs is described by the same Mittag–Leffler distribution~\cite{blumenfeld1997levy}, indicating robust universality largely insensitive to microscopic details.

\textit{Continuous-Time Random Walk.}
The CTRW describes a minimal trapping-jumping dynamics in which a particle performs independent jumps $\delta x$ drawn from a PDF $f(\delta x)$, separated by waiting times $\tau$ drawn from $\psi(\tau)$. 
Originally introduced to model transport in amorphous materials~\cite{scher1975anomalous}, it is widely applied to systems exhibiting anomalous diffusion~\cite{kutner2017continuous}. 
We use CTRW to introduce a separation between physical time and an internal clock, a distinction we later extend to other transport models. 
We focus on power-law waiting times, $\psi(\tau)\sim A \tau^{-1-\alpha}$ for large $\tau$ with $0<\alpha<1$, such that the mean waiting time diverges, $\int_{0}^{\infty}\tau\,\psi(\tau)\,d\tau=\infty$, and we assume a finite jump variance, $\int d(\delta x)\,(\delta x)^2 f(\delta x)=a^2$.

First, we highlight the dual role of time in CTRW. 
The physical time $t$ is fixed by the observer, while the number of jumps $N$ completed within this interval varies across trajectories due to random waiting times. 
Conversely, if $N$ is fixed, then $t=\sum_{i=1}^{N}\tau_i$ is random. 
Since the $\{ \tau_i \}$ are independent identically distributed random variables drawn from a heavy-tailed $\psi(\tau)$, their sum converges for large $N$ to a one-sided L\'{e}vy-stable law~\cite{bouchaud1990anomalous}, yielding the scaling
\begin{equation}
    \label{eq:ctrwNtot}
    t\sim N^{1/\alpha}\eta.
\end{equation}
Here $\eta$ is a positive random variable with L\'{e}vy-stable density $l_{\alpha,A,1}(\eta)$, defined via its Laplace transform as
$\int_0^\infty e^{-\eta s} l_{\alpha,A,1}(\eta)\,d\eta = e^{-As^{\alpha}}$, $(A>0)$. 
Although this scaling is obtained by fixing $N$ and viewing $t$ as random, it can be interpreted as the asymptotic stochastic relation between $t$, $N$, and $\eta$ along trajectories. 
Inverting it at fixed $t$ and treating $N$ as random, gives $N=(t/\eta)^\alpha$, so the distribution $Q_t(N)$ of the number of jumps $N$ performed up to time $t$ follows by a change of variables applied to $\eta$. 
This motivates the subordination approach: one first computes an observable at fixed $N$ and then averages over $N$ using $Q_t(N)$. 
For example, the probability density to find the particle at position $x$ at time $t$, $P(x,t)$, is 
$P(x,t)=\sum_{N=0}^{\infty}\mathcal P_N(x)\,Q_t(N)$, where $\mathcal P_N(x)$ is the displacement distribution after exactly $N$ jumps. 
Thus, $Q_t(N)$ establishes the connection between the internal clock $N$ and the physical time $t$.

\begin{figure}[ht!]
\centering
\includegraphics[width=0.45\textwidth]{"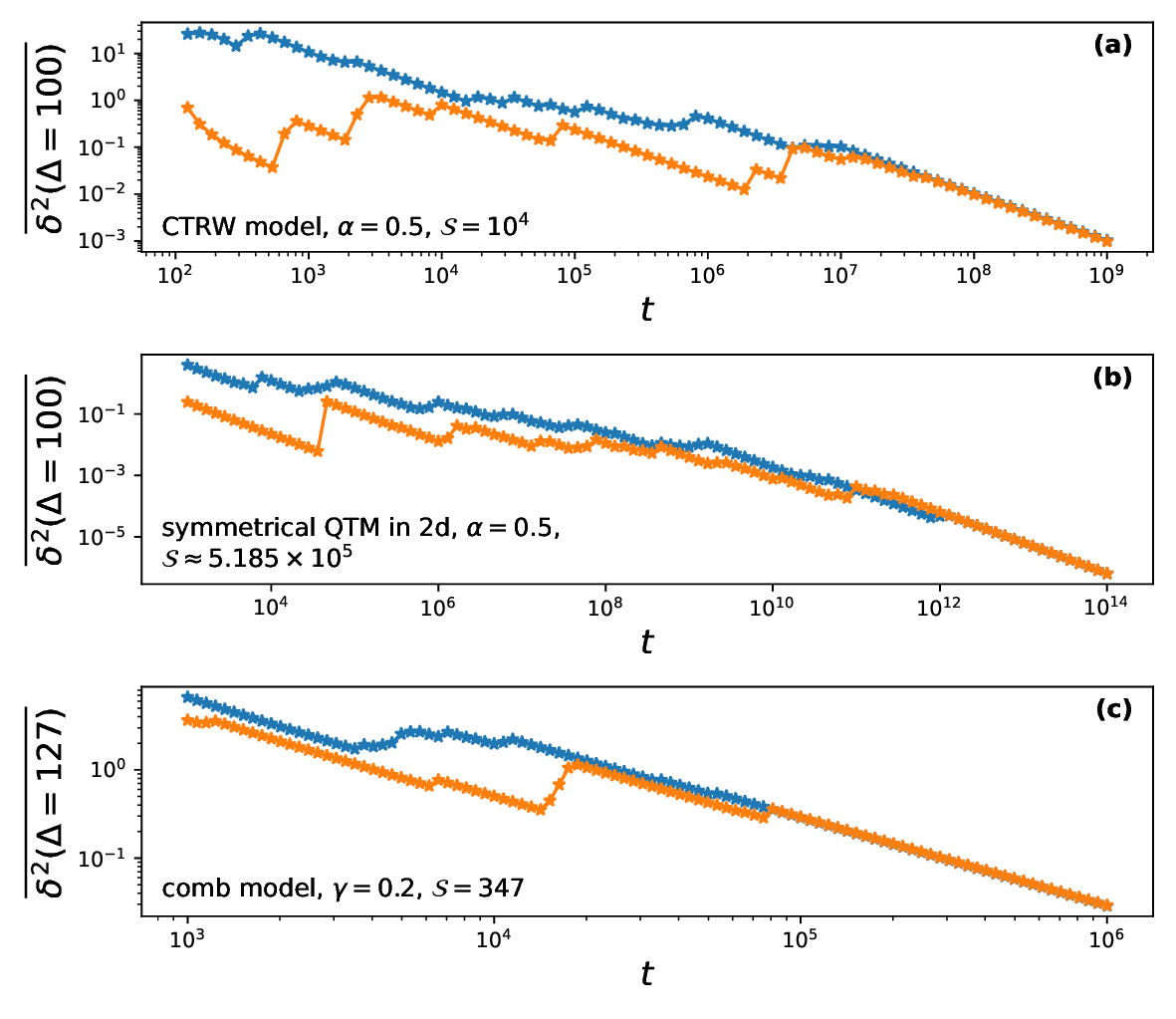"}
\vspace{-12pt}
\caption{ \textit{Conditional ergodicity.} 
In each model—(a) CTRW, (b) QTM, and (c) comb model —two independent TA-MSD trajectories are plotted as functions of the physical time $t$. Both trajectories are conditioned to reach the same internal clock value $\mathcal{S}$ at the end of the measurement. In all cases, the two TAs overlap well before the final time, illustrating conditional ergodicity [Eq.~\eqref{eq:conditional_ergodicity}].
}
\label{Fig:2}
\end{figure}

This approach, of first examining behavior as a function of the internal clock and then transforming the resulting behavior to physical time, is applied to the TA-MSD $\overline{\delta^2}(\Delta, t)$. 
First, examine $\overline{\delta^2(\Delta,t)}$ conditioned on a fixed internal clock, i.e., a fixed $N$. 
Conditioning defines $\overline{\delta^2(\Delta,t)}|_N$ as the TA-MSD evaluated over trajectories that have completed exactly $N$ jumps by time $t$. 
For large fixed $N$, the integrand $\left[x(t'+\Delta)-x(t')\right]^2$ in Eq.~\eqref{eq:delta_definition} is sampled across many states and a broad range of values along the trajectory, so that TAs converge to EAs when conditioned on the internal clock $N$, a property we refer to as conditional ergodicity.
In this regime, $\left[x(t'+\Delta)-x(t')\right]^2$ is set by the number of jumps occurring within lag-time $\Delta$, times $a^2$, the variance of a single jump. 
For $\Delta \ll t$, an interval of length $\Delta$ within $[0,t]$ contains on average $N\Delta/t$ jumps, yielding
\begin{equation}
    \label{eq:ctrwlineargrowth}
    \overline{\delta^2(\Delta, t)}|_N \sim a^2 \Delta\frac{ N}{t}.
\end{equation}
This linear dependence on $\Delta$ is evident in Fig.~\ref{Fig:1}(a), which shows the TA-MSD for individual numerically simulated trajectories. 
The distinct curves correspond to different values of $N$.
Conditional ergodicity is directly visible in Fig.~\ref{Fig:2}(a). 
For fixed $\Delta$, the TA-MSD $\overline{\delta^2(\Delta,t)}|_N$ shows trajectory-dependent fluctuations at short measurement times $t$, but for two trajectories matched to the same final internal clock $N$ these fluctuations collapse at long $t$, and the two traces become indistinguishable.

The conditioned TA-MSD, $\overline{\delta^2(\Delta,t)}|_N$, constitutes the first step of the subordination approach. 
The remaining step is to average over the distribution $Q_t(N)$ at fixed physical time $t$. 
Specifically, since $N=(t/\eta)^{\alpha}$, a random-variable transformation from $\eta$ to $N$
yields
\begin{equation}
    \label{eq:qtnCTRW}
  Q_t(N)\sim \frac{t}{\alpha N^{1+1/\alpha}}\, l_{\alpha}\!\left(t/N^{1/\alpha}\right)   .
\end{equation}
The rescaled variable, $\xi \equiv \overline{\delta^2(\Delta,t)}/\langle \overline{\delta^2(\Delta,t)}\rangle$, captures the scatter of the TA-MSD
According to Eq.~\eqref{eq:ctrwlineargrowth}, $\xi=N/\langle N\rangle$, and Eq.~\eqref{eq:qtnCTRW} yields the PDF of $\xi$, 
\begin{equation}
    \label{eq:MittagLefflerDistribution}
    \phi_\alpha(\xi)=\frac{\Gamma^{1 / \alpha}(1+\alpha)}{\alpha \xi^{1+1 / \alpha}}\, 
    l_\alpha\!\left[\frac{\Gamma^{1 / \alpha}(1+\alpha)}{\xi^{1 / \alpha}}\right]
\end{equation}
which depends only on $\alpha$. The distribution $\phi_\alpha(\xi)$ is known as the Mittag-Leffler distribution~\cite{blumenfeld1997levy}. 

\begin{figure}[ht]
\centering
\includegraphics[width=0.45\textwidth]{"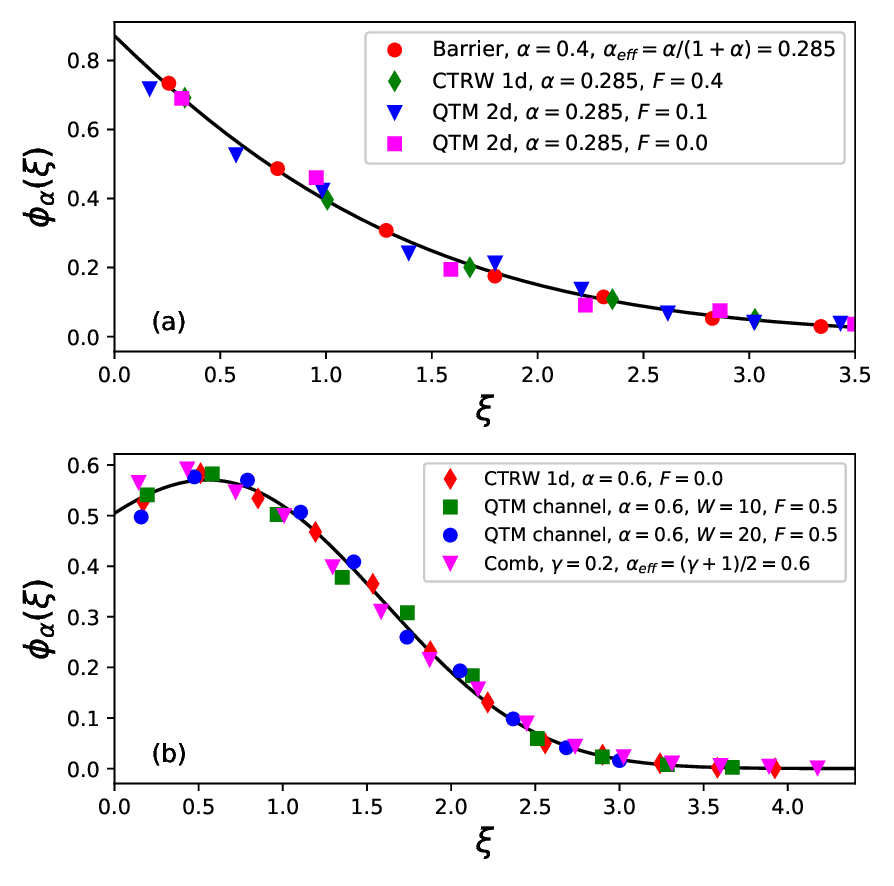"}
\vspace{-15pt}
\caption{\textit{The universal distribution across all models.} The PDF $\phi_\alpha (\xi)$ [in Eq.~\eqref{eq:MittagLefflerDistribution}] capturing the scatter of the rescaled TA observable $\xi=\overline{\mathcal {O} }/\langle \overline{\mathcal {O}}\rangle$ across different models and under varying geometric constraints and external forces (see SM~\cite{SM}). Once $\alpha$ associated with the trapping times is fixed, all models collapse to the same Mittag-Leffler form without any fitting. 
Panel (a): $\alpha=0.285$; the scatter of $\overline{\mathcal {O}}=|\overline{\delta}|$ in the Barrier model, $\overline{\mathcal {O}} = \overline{\delta}$ in biased $1d$ CTRW and transient $2d$ QTM, and of $\overline{\mathcal {O}} = \overline{\delta^2}$ in the symmetrical $2d$ QTM .
Panel (b): $\alpha=0.6$; the scatter of $\overline{\mathcal {O}} = \overline{\delta^2}$ in symmetrical $1d$ CTRW, $\overline{\mathcal {O}} = \overline{\delta}$ in biased QTM under geometrical constraints of a channel for two cases of width $W=10,20$ and of $\overline{\mathcal {O}} = \overline{\delta^2}$ in the Comb model.
}
\label{Fig:3}
\end{figure}

The existence of scatter in the TA-MSD for CTRW, and its Mittag-Leffler form, was established in Ref.~\cite{he2008random}. 
The central question we address is whether this behavior survives beyond renewal CTRW dynamics in settings where quenched disorder, correlations, geometric constraints, or external forces violate the CTRW  assumptions. 
Figures~\ref{Fig:1} and \ref{Fig:2} show that two key features
persist in these models: (i) TA transport coefficients display pronounced trajectory-to-trajectory scatter (Fig.~\ref{Fig:1}), 
(ii) Conditional ergodicity holds when conditioning on a natural internal clock, restoring convergence at large $t$ (Fig.~\ref{Fig:2}). 
Our main finding is that this combination leads to universal fluctuations: for a broad class of observables $\mathcal O$, 
the distribution of the rescaled TA (divided by its EA)$, \xi=\overline{\mathcal {O} (\Delta)}/\langle \overline{\mathcal {O}(\Delta)}\rangle$, collapses
onto the same Mittag-Leffler form, as demonstrated in Fig.~\ref{Fig:3}. 
This motivates identifying when conditional ergodicity imposes Mittag-Leffler fluctuations.

From the CTRW derivation we extract four ingredients that control the scatter of TA transport coefficients: (i) the existence of an internal clock $N$; 
(ii) a L\'{e}vy-stable mapping between physical time and the internal clock, $t\sim N^{1/\alpha}\eta$ [Eq.~\eqref{eq:ctrwNtot}], inherited from heavy-tailed sojourn statistics; 
(iii) conditional ergodicity 
\begin{eqnarray} \label{eq:conditional_ergodicity}
    P\!\left(\overline{\mathcal{O}(t)}|_{\mathcal{S}}\right)
\xrightarrow[t\to\infty]{}
\delta\!\left(\overline{\mathcal{O}(t)}|_{\mathcal{S}}-\langle \overline{\mathcal {O}(t)}|_{\mathcal{S}} \rangle\right),
\end{eqnarray}
in the large $\mathcal{S}$ limit, while $P(\cdots)$ is the PDF of $\overline{\mathcal {O}(t)}$ and $\delta(\cdots)$ is the Dirac $\delta$-function.
(iv) an asymptotically linear dependence of the conditioned TA on the clock, $\overline{\delta^2}|_N\propto N$ [Eq.~\eqref{eq:ctrwlineargrowth}].

This reasoning extends beyond CTRW: if a given transport model admits an internal clock ${\cal S}$ and a TA observable $\overline{\cal O}$ satisfying the same four ingredients, then the CTRW derivation carries through with the replacements $N\!\to\!{\cal S}$ and TA-MSD$\!\to\!\overline{\cal O}$, yielding universal Mittag-Leffler scatter for the normalized TA.
We continue to examine other models that exhibit the scatter described in Fig.~\ref{Fig:1}.

\textit{Quenched trap model.} 
The quenched trap model (QTM) describes transport in a disordered medium through site-dependent energetic traps~\cite{bouchaud1990anomalous}. 
Each lattice site ${\bf r}$ carries a quenched trap depth $E_{\bf r}$, leading to an activated escape with an Arrhenius law for the escape time $\tau_{\bf r}\propto e^{E_{\bf r}/T}$ ($k_B=1$). 
For exponentially distributed trap depths with mean $T_g$, the induced waiting-time statistics are heavy tailed, $\psi(\tau)\sim \tau^{-1-\alpha}$ with $\alpha=T/T_g<1$. 
Unlike CTRW, the quenched landscape generates strong correlations (through repeated visits to the same sites), which leads to a distinct response to external fields, and geometry dependence~\cite{bouchaud1990anomalous,monthus1996models,monthus2004nonlinear,burov2011time,burov2012weak,burov2017quenched,burov2020transient,akimoto2020trace,shafir2022case,shafir2024disorder}.  
Nevertheless, the internal-clock principle carries over. 
In QTM the internal clock is ${\cal S}=\sum_{\bf r} n_{\bf r}^{\alpha}$ (see SM~\cite{SM}, Sec.~I), where $n_{\bf r}$ counts visits to site ${\bf r}$~\cite{burov2011time}, and the physical time obeys the same L\'{e}vy-stable mapping $t\sim {\cal S}^{1/\alpha}\eta$~\cite{burov2011time}. 
Conditional ergodicity is visible in Fig.~\ref{Fig:2}(b): at fixed ${\cal S}$, the TA observables converge to the same value.
Finally, once conditional ergodicity holds, the relevant TA observables grow linearly with ${\cal S}$ (see SM~\cite{SM}, Eq.~(S.9)), which ensures that the scatter PDF $\phi_\alpha(\xi)$ is inherited from the L\'{e}vy-stable time change. 
This applies not only to the TA-MSD in a symmetrical process but also to the time-averaged displacement in the biased case, $\overline{\delta(\Delta,t)} \equiv (t-\Delta)^{-1}\int_{0}^{t-\Delta} [x(t'+\Delta)-x(t')]\,dt'$.
The trajectory-to-trajectory scatter is shown in Fig.~\ref{Fig:1}(b), and the normalized variable $\xi$ (capturing it) collapses onto the same Mittag-Leffler form for all the different scenarios depicted in Fig.~\ref{Fig:3} - different geometries, with and without an external force, and across temperatures.

\textit{Comb model.} 
The comb model produces anomalous transport through geometric constraints: motion along the backbone is intermittently slowed down by excursions into dead-end ``teeth'' that ultimately return the particle to the backbone~\cite{iomin2018fractional,sandev2016comb,illien2016propagators,weiss1986some,havlin1986diffusion,ben2000diffusion}. 
We consider a quenched comb with finite tooth lengths $L$ drawn from a heavy-tailed distribution $p(L)\sim L^{-1-\gamma}$ for $L\gg 1$ with $0<\gamma<1$, relevant to self-similar geometries such as critical percolation clusters~\cite{bunde1986diffusion,havlin1986diffusion,ben2000diffusion}. 
The waiting time between successive backbone steps is set by the return time from a tooth: for a fixed tooth length $L$, the exit-time distribution obeys $\psi(\tau|L)\sim \tau^{-3/2}e^{-\tau/L^2}$~\cite{bunde1986diffusion,redner2001guide}. 
Averaging over the quenched distribution of $L$ yields an effective heavy-tailed waiting-time distribution $\psi(\tau)\sim \tau^{-1-\alpha_{\rm eff}}$ with $\alpha_{\rm eff}=(1+\gamma)/2$. 
In this model, the internal clock is the number of steps along the backbone, ${\cal S}=N_x$, and conditional ergodicity is visible in Fig.~\ref{Fig:2}(c).
The scatter $\phi(\xi)$ of the normalized TA-MSD collapses onto the Mittag--Leffler form of Eq.~\eqref{eq:MittagLefflerDistribution} with $\alpha=\alpha_{\rm eff}$, as shown in Fig.~\ref{Fig:3}.

\textit{Barrier model.} 
The random barrier model is a random walk on a lattice in which each link is associated with a quenched barrier height, and the time to cross a given link follows an Arrhenius law~\cite{alexander1981excitation}. 
For exponentially distributed barrier heights, the corresponding crossing times are scale free, $\psi(\tau)\sim \tau^{-1-\alpha}$. 
Because the same crossing time applies in both directions across a given link, the dynamics are strongly correlated in space and time.
Locally, the transition rates are asymmetric, giving the particle a preferred direction of motion, yet it typically becomes trapped within an effective ``cell'' bounded by two significantly large barriers. 
Analytical results for the barrier model are scarce, and no explicit internal clock is known.
Yet, using Alexander's result~\cite{alexander1981excitation}, the escape time from an effective cell decays as $\tau^{-\alpha/(1+\alpha)}$, implying an effective exponent $\alpha_{\rm eff}=\alpha/(1+\alpha)$ characterizing the sojourn statistics in these cells.

To examine scatter beyond the TA-MSD, we consider the absolute TA displacement $|\overline{\delta(\Delta,t)}|=|(t-\Delta)^{-1}\int_0^{t-\Delta}x(t'+\Delta)-x(t')\,dt'|$, 
whose variability is illustrated in Fig.~\ref{Fig:1}(d). 
An explicit form of ${\cal S}$ has not been derived for the barrier model, and we use it as a numerical test of whether the normalized scatter follows the Mittag-Leffler form predicted by the internal-clock scenario. 
Figure~\ref{Fig:3} shows that the normalized scatter $\phi(\xi)$ is described by a Mittag-Leffler form, consistent with applying the internal-clock scenario with the effective exponent $\alpha_{\rm eff}$. 
Additional results are provided in Sec.~II of the SM~\cite{SM}.

\textit{Discussion.} The most striking result of this work is the collapse shown in Fig.~\ref{Fig:3}: despite different microscopic mechanisms, the rescaled variable $\xi$, which quantifies the trajectory-to-trajectory scatter of time-averaged observables (e.g., the TA-MSD), follows the same distribution [Eq.~\eqref{eq:MittagLefflerDistribution}] across all models.
This collapse shows that the variability is not model-specific but reflects an intrinsic statistical property.
The Mittag–Leffler curves in Fig.~\ref{Fig:3} involve no fitting parameters; once the trapping exponent $\alpha$ is fixed, the resulting distribution is fully determined.
Moreover, universality holds for different geometric constraints, dimensionality, and various TA observables.

Mittag-Leffler distribution has been reported in other systems, including 
weakly chaotic deterministic maps~\cite{akimoto2010role,akimoto2012aging,korabel2013distribution} and stochastic dynamics with quenched disorder~\cite{akimoto2016universal}.
We present a unified framework that explains why this universality emerges, based on two ingredients: an internal clock ${\cal S}$ and conditional ergodicity [Eq.~\eqref{eq:conditional_ergodicity}]. 
In WEB, TAs taken over a measurement time window $t$ remain trajectory dependent and therefore do not self-average in the usual sense. 
We show that this variability is controlled by fluctuations of the internal clock ${\cal S}$: conditioned on ${\cal S}$, the trajectory-to-trajectory scatter of TAs collapses.
Thus, the dynamics are ergodic when conditioned on ${\cal S}$, even though they are non-ergodic in physical time $t$.

This separation partitions WEB into two parts: (i) a conditionally ergodic component at fixed ${\cal S}$, and (ii) randomness originating from the mapping between physical time $t$ and the internal clock. 
For systems with power-law sojourn statistics, this mapping is governed by a L\'{e}vy-stable law [Eq.\eqref{eq:ctrwNtot}].
Once normalized by their mean, TAs inherit their statistics from this time change, yielding the Mittag–Leffler distribution for the corresponding normalized transport coefficients. 
Importantly, across the different models we studied, the internal clock need not be a simple renewal counter. From this perspective, the collapse in Fig.~\ref{Fig:3} is a fingerprint of a shared L\'{e}vy-stable time-change structure across the different models.

A final point concerns when Mittag–Leffler statistics should be expected. In our framework, the Mittag–Leffler law arises when the TA observable is conditionally ergodic and, at fixed internal clock ${\cal S}$, grows asymptotically linearly with ${\cal S}$. 
In that case, rescaling by the ensemble average produces fluctuations governed by the L\'{e}vy-stable time change, resulting in a Mittag–Leffler distribution.
For other asymptotic dependencies on ${\cal S}$, the scatter is still L\'{e}vy-controlled but follows the corresponding transformed law of $\eta$, rather than Mittag–Leffler distribution.

\textit{Data availability}---The data that support the findings of this article are openly available~\cite{shafirGithub}.

\begin{acknowledgments}
{\emph{Acknowledgments.}} This work was supported by the Israel Science Foundation, Grant No.~3791/25.
\end{acknowledgments}

\bibliography{./main.bib}

\appendix

\clearpage

\onecolumngrid

\begin{center}
{\large \textbf{Supplemental Material for: Conditional Ergodicity and Universal Fluctuations in Weak Ergodicity Breaking}}
\end{center}

\setcounter{secnumdepth}{3}
\setcounter{section}{0}
\renewcommand{\thesection}{\Roman{section}}

\setcounter{equation}{0}
\renewcommand{\theequation}{S\arabic{equation}}

\vspace{0.5cm}

Supplemental material includes 
(\ref{SM: one}) The QTM model derivations, including the internal clock, geometrical constraints, and response to an external force. 
(\ref{SM: two}) Additional Numerical results and discussion of the Barrier model that include simulations of the TA-MSD.
(\ref{SM: three}) Simulations procedures explaining the methods used to simulate trajectories across the different models.

\vspace{0.5cm}

\twocolumngrid

\section{Fluctuations in the QTM follow a Mittag-Leffler distribution \label{SM: one}}
In this section we show analytically that the transient QTM obeys the same universal Mittag-Leffer distribution, that which appears in the main text in Eq.~(6).

We start by showing that the internal clock can be defined in the QTM by $S=\sum_{\mathbf{r}}\left(n_{\mathbf{r}}(t)\right)^{\alpha}$ where $n_{\mathbf{r}}$ is the number of visitations to each site. 
The internal clock of the QTM plays the same role as the number of jumps $N$ in the CTRW.
In particular we follow the proof in in Ref. \cite{burov2020transient} and show that the random variable $\eta = t/(S)^{1/\alpha}$ has the PDF, $l_\alpha(\eta)$, in the $t \to \infty$ and $S \to \infty$ limit. 

The total time in the QTM is $t=\sum_{\mathbf{r}} n_{\mathbf{r}} \tau_{\mathbf{r}}$ with $n_{\mathbf{r}}$ the amount of visitations, and $\tau_{\mathbf{r}}$ the fixed waiting time of a lattice site $\mathbf{r}$.
By fixing the values of $\{ n_{\mathbf{r}} \}$ and  averaging over $\{\tau_{\mathbf{r}}\}$ (disorder averaging) the $t \to u$ Laplace transform of the PDF of $t$ is
\begin{equation}
    \langle e^{-u t} \rangle =\displaystyle \langle \exp\left( -u \sum_{\bf r} n_{\bf r}\tau_{\bf r}\right) \rangle.
    \label{etalaplace}
\end{equation}
Since the $\{\tau_{\bf r}\}$ are i.i.d. with PDF $\psi(\tau_{\mathbf{r}}) \sim A \tau^{-1-\alpha}$ we have
$\langle e^{-u t} \rangle =\displaystyle \prod_{\bf r} {\hat{\psi}}\left[n_{\bf r}u\right]$
where $\hat{\psi}(u)$ is the Laplace transform of a single $\tau_{\mathbf{r}}$, ${\hat{\psi}}(u)=\int_0^\infty\exp(-\tau_{\bf r} u)\psi(\tau_{\bf r})\,d\tau_{\bf r}$.
In the small $u\to 0$ limit, ${\hat{\psi}}(u)\sim 1-Au^\alpha$, and we have
\begin{equation}
\langle e^{-u t} \rangle =\displaystyle \prod_{\bf r} \left( 1-A {n_{\bf r}^\alpha}u^\alpha\right) \sim e^{-A Su^{\alpha}},
\end{equation}
by taking the leading term in the multiplication.
The Laplace transform of the one-sided L\'evy distribution $l_{\alpha}(\eta)$ is $\sim e^{-A u^{\alpha}}$. Therefore, in the long time limit ($t\to\infty$), $t / S^{1/\alpha} \sim \eta$, where $\eta$ is a random variable distributed according to $l_{\alpha}(\eta)$.
We can now invert the process measurement time $t$ to find the PDF of $S$
\begin{equation} \label{eq:S_alpha_PDF}
\mathcal{P}_{t}\left(S\right) \sim \frac{t}{\alpha} S^{-1 / \alpha-1} l_{\alpha}\left(\frac{t}{S^{1 / \alpha}}\right) \quad(t \rightarrow \infty).
\end{equation}
Note that this PDF is identical to that of a CTRW upon identifying $S$ as the internal clock rather than $N$ (see Eq.~(5) in the main text).

Obtaining the internal clock for the QTM allows us to find the distribution of $\xi$ in the QTM by utilizing a subordination approach defined in Ref.~\cite{burov2020transient}. Below, we outline the procedure.

We consider the QTM in dimension $d=2$ with a constant force $F$ applied along the $x$ direction, pulling on the random walker making the process nonrecurrent. 
This results in a mean jump size $\langle \delta x \rangle>0$ in the $x$-direction and $\langle \delta y \rangle= 0$ in the $y$-direction.
The single jump probabilities obey $p_\rightarrow / p_\leftarrow = \exp (a F / k_B T)$ and $p_\uparrow  / p_\downarrow = 1$ as dictated by detailed balance, where $k_B$ is Boltzmann's constant and $T$ is the temperature. 
Therefore, $\langle \delta x \rangle = a \tanh (a F/4 k_B T)$. 
For the transient QTM, similar arguments to those in the main text show that the TA-displacement in the $x$-direction (parallel to the applied external force $F$),
\begin{eqnarray} 
    \overline{\delta (\Delta, t)} = \frac{1}{t-\Delta} \int_0^{t-\Delta}[x(t'+\Delta)-x(t')] \mathrm{d} t',
\end{eqnarray}
has the asymptotic behavior 
\begin{eqnarray} \label{eq:TA_displacment_QTM}
    \overline{\delta(\Delta, t)}  \sim \Delta \langle \delta x\rangle N / t, \quad t \to \infty,
\end{eqnarray}
with $\langle \delta x \rangle = a \tanh (a F / 4 k_B T) > 0$.
This follows from realizing that the integrand has the same distribution of sojourn time in state $\left[x\left(t^{\prime}+\Delta\right)-x\left(t^{\prime}\right)\right]=0$ as the original process $x(t)$ with a waiting time PDF $\psi(\tau) \sim \tau^{-(1+\alpha)}$.
Finding the distribution of $N$ will now determine the statistics of $\overline{\delta}$ for a fixed $t$.
Using subordination~\cite{klafter2011first, weiss1994aspects} on $S$, we can find the PDF of $N$ from the result in Eq.~\eqref{eq:S_alpha_PDF},
\begin{eqnarray} \label{eq:pdf_N_QTM_sum}
    Q_t^*(N) = \sum_{S} \mathcal{G}_{S, \mathbf{r}}(N) \mathcal{P}_{t}\left(S\right) ,
\end{eqnarray}
which stems from the law of total probability. 
Here, $\mathcal{G}_{S, \mathbf{r}}(N)$ is the probability of different values of $N$ for a prescribed $S$ and $\mathbf{r}$, and the star notation is to differentiate the result in the QTM from the one in CTRW.
We determine $Q_t^*(N)$ by using the result 
\begin{equation} \label{eq: limit G}
\mathcal{G}_{S}(N, \mathbf{r}) \underset{S \rightarrow \infty} \longrightarrow \delta\left(S-\Lambda N\right).
\end{equation}
in the transient regime, here, $\delta$, on the right hand side, is Dirac's delta function. 
The full proof appears in Ref.~\cite{burov2020transient} and relies on calculating the mean $S$ and the variance $ \text{Var}( S) \rightarrow 0 $ in the long time (equivalently large $N$) limit using generating functions formalism.
The local time $S$ and the number of steps $N$ are found to have a linear connection $S \to \Lambda N$ in the long time limit with
\begin{eqnarray} \label{eq:Lambda definition}
    \Lambda=\left[\left(1-Q_0\right)^2 / Q_0\right] L i_{-\alpha}\left(Q_0\right).
\end{eqnarray}
Here, $L i_a(b)=\sum_{j=1}^{\infty} b^j / j^a$ is the polylogarithm function~\cite{gradshteyn2014table} and $Q_0$ is the probability of the random walker to eventually return to the origin~\cite{weiss1994aspects} which must be $<1$ for the connection to be valid.
This means that Eq.~\eqref{eq:TA_displacment_QTM} turnes into
\begin{equation} \label{eq:appx-qtm}
    \overline{\delta(\Delta, t)} \sim \Delta \langle \delta x \rangle (S / \Lambda) t,
\end{equation}
establishing a linear dependence on the internal clock $S$ as expected from the four main ingredients discussed in the main text that lead to universal fluctuations.
Utilizing Eq.~\eqref{eq: limit G} in Eq.~\eqref{eq:pdf_N_QTM_sum} and turning the sum into an integral, as usually done in subordination~\cite{weiss1994aspects, klafter2011first}, yields
\begin{equation} \label{eq:N distribution}
Q^*_t(N) \sim \frac{t / \Lambda^{1 / \alpha}}{\alpha N^{\frac{1}{\alpha}+1}} l_{\alpha, A, 1}\left(\frac{t / \Lambda^{1 / \alpha}}{N^{1 / \alpha}}\right).
\end{equation}
Eq.~\eqref{eq:N distribution} allows us to find the ensemble averaged $\langle \overline{ \delta (\Delta , t)}\rangle$ using Eq.~\eqref{eq:TA_displacment_QTM},
\begin{eqnarray} \label{eq:SM_TA_MSD_quenched}
    \langle\overline{\delta(\Delta, t)}\rangle \sim \Delta \left\langle\delta x\right\rangle / \left[{A \Lambda \Gamma(1+\alpha)} t^{1-\alpha}\right].
\end{eqnarray}
Here, $\Lambda$ quantifies the difference between quenched and annealed disorder (the CTRW) as a function of the return probability $Q_0$ - which depends on the geometry (type of lattice, dimensions, etc.) and any external forces present in the system.
By plugging in a specific systems $Q_0$, that of a channel geometry~\cite{shafir2024disorder} for instance, we are able to find the ensemble averaged TA-MSD $\langle \overline{\delta} \rangle$.
We now obtain the PDF of $\xi = \overline{\delta} / \langle \overline{\delta}\rangle$ by change of variables using Eqs.~\eqref{eq:TA_displacment_QTM} and \eqref{eq:N distribution},
\begin{equation} \label{eq:xi PDF QTM}
\lim _{t \rightarrow \infty} \phi_\alpha(\xi)=\frac{\Gamma^{1 / \alpha}(1+\alpha)}{\alpha \xi^{1+1 / \alpha}} l_\alpha\left[\frac{\Gamma^{1 / \alpha}(1+\alpha)}{\xi^{1 / \alpha}}\right]
\end{equation}
which is exactly the solution obtained for the CTRW~\cite{he2008random}.
While his result was developed for the case of transient QTM, simulations show that even for the recurrent case ($F=0$) in dimensions $d \le 2$, where exact analytical results are still unknown, the Mittag-Leffler distribution in Eq.~\eqref{eq:xi PDF QTM} still applies.

\section{Additional Numerical results and discussion for the Barrier model \label{SM: two}}
We now discuss in greater detail the barrier model and its implications for ergodicity. As presented in the main text, the barrier model is defined as a random walk on a lattice in which each link acts as a symmetric barrier~\cite{alexander1981excitation}. We focus on the case of scale-free waiting times for crossing a barrier; specifically, for $\tau \gg 1$, the waiting-time PDF follows $\psi(\tau) \sim \tau^{-1-\alpha}$.

\begin{figure}[ht!]
\centering
\includegraphics[width=0.45\textwidth]{"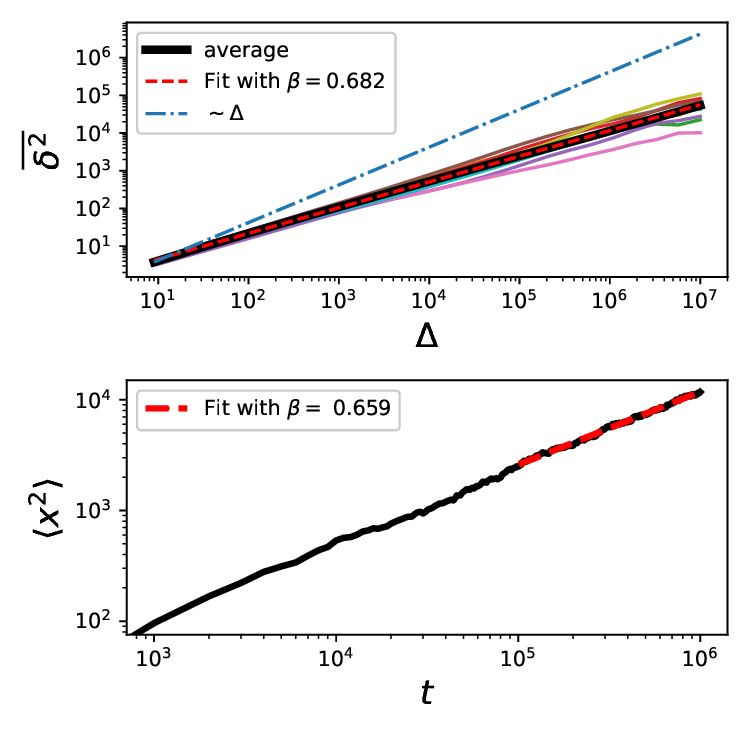"}
\caption{Simulation results of $\overline{\delta^2}$ for the Barrier model with $\alpha=0.5$, $T=10^8$ showing ergodic behavior. Upper plot: The ensemble average of the TA-MSD $\langle \overline{\delta^2} \rangle$ (solid black line) is computed over $1000$ trajectories. Ten individual realizations of $\overline{\delta^2}$ are also shown, color-coded. Lower plot: The corresponding EA-MSD. The fitting exponent $\beta$ in both plots agrees very well with the expected theoretical value~\cite{bouchaud1990anomalous, alexander1981excitation}, $\beta \approx 2 \alpha_{\text{eff}} = 2 \alpha / (1+\alpha) = 2/3$.
}
\label{Fig:appendix-2}
\end{figure}

With this setup in mind, we turn to the TA-MSD in the barrier model. In contrast to the CTRW~\cite{he2008random, lubelski2008nonergodicity}, where the TA-MSD scales linearly with the lag time, $\sim \Delta$, numerical investigations indicate that the barrier model is ergodic, at least with respect to the second moment of the displacement. In particular, we find
\begin{equation}
\langle x^2(t) \rangle \propto t^{\frac{2\alpha}{1+\alpha}},
\qquad
\langle \overline{\delta^2(\Delta)} \rangle \propto \Delta^{\frac{2\alpha}{1+\alpha}},
\end{equation}
see Fig.~\ref{Fig:appendix-2}. The scaling of the ensemble-averaged MSD (EA-MSD) is consistent with Refs.~\cite{derrida1988introduction, bouchaud1990anomalous, alexander1981excitation}.

\begin{figure}[ht!]
\centering
\includegraphics[width=0.45\textwidth]{"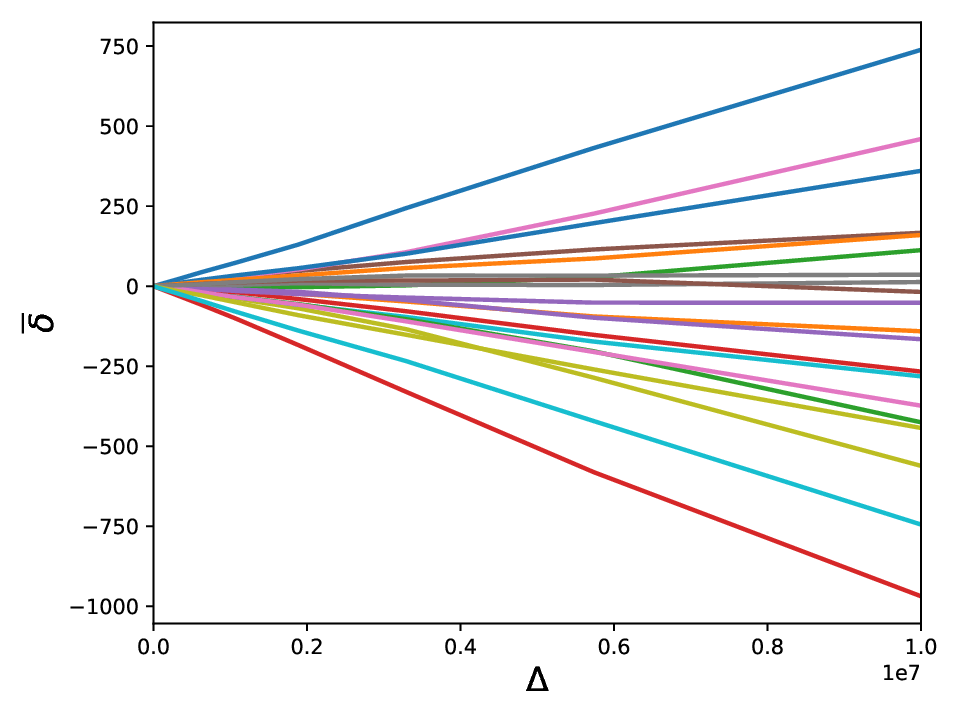"}
\caption{Simulation results of $\overline{\delta}$ for the Barrier model with $\alpha=0.8$, $T=10^8$ showing 20 independent examples color coded.
}
\label{Fig:appendix-3}
\end{figure}

Our hypothesis for this result in the barrier model is the following: the quenched, scale-free transition probabilities cause the particle to follow a path of least resistance until it gets stuck between two very high barriers. 
Eventually, a rare event will cause the particle to escape this enclosed region, defining a new effective cell size. 
The particle will prefer not to jump backwards (through an extremely high barrier), causing a net drift until it encounters a new barrier of equal or larger height (this is always possible).
Once the particle overcomes this second barrier, it becomes even less probable for the particle to return to the origin, as it needs to jump against two very high barriers.
The process repeats, with three, four, and more such high barriers spaced at effective cell size $\chi$, causing a biased motion with a random $\overline{\delta} \sim v_\alpha \Delta$. 
Applying Alexander's~\cite {alexander1981excitation} result, the decay rate of the probability to be at the origin at time $t$ is $P_0(t) \sim t^{-\alpha / (\alpha +1)}$.
This indicates an escape time $\tau$ out of an effective cell size $\chi$ with the PDF $\psi(\tau) \sim \tau^{-1-\alpha / (\alpha + 1)}$ in analogy to the CTRW.
Therefore, the total time is $t \sim \sum_{i=1}^{S} \tau_i$ with $S$ the number of effective cells jumped up to time $t$, which we hypothesize to be the internal clock of the model.
Notice that we neglect repeated samples of the $\{ \tau_i \}$ as we assume a directed motion, as evident by the scattering of $\overline{\delta}$ (instead of $\overline{\delta^2}$) in Fig.~\ref{Fig:appendix-3}.
A directed motion assumption also fits with the EA result $\langle x^2 \rangle \sim v^2 t^{2 \alpha / (1 + \alpha)}$ with unknown $v$, obtained in a different manner in Ref.~\cite{bouchaud1990anomalous}.
Applying L\'evy's limit theorem, we have $\sum_{i=1}^{N_\chi} \tau / (N_\chi)^{1/\alpha_{\text{eff}}} \sim \eta$ with $\eta$ distributed according to a L\'evy PDF, $l_{\alpha_{\text{eff}}}(\eta)$, and $\alpha_{\text{eff}} = \alpha / (\alpha +1)$.
This is precisely what we see in Fig.~3 in the main text, the scattering of $\xi = |\overline{\delta}| / \langle |\overline{\delta}| \rangle$ obeys the Mittag-Leffler form of $\Phi_{\alpha_{\text{eff}}}(\xi)$ [Eq.~(6) in the main text]. 
The only requirement was to find the appropriate $\alpha_{\text{eff}}$.

\section{Simulation procedures \label{SM: three}}
In this section we discuss how we simulated the trajectories in the various models.

As presented in the main text, the barrier model is defined as a random walk on a lattice where each link acts as a symmetrical barrier~\cite{alexander1981excitation}. 
The sojourn time $\tau$ to jump over a barrier depends exponentially on the height, as dictated by Arrhenius' law, $\tau_{i, i+1} \propto \exp (E_{i , i+1} / T)$ where $E_{i , i+1}$ is the energy of the link between sites $i$ and $i+1$ and we have $E_{i+1 , i} = E_{i , i+1}$. 
When the energies are exponentially distributed, $\phi(E) \sim \exp (-E / T_g) / T_g$ where $T_g$ is the characteristic temperture, we receive 
\begin{equation}
    \psi(\tau_{i, i+1}) = \phi(E_{i, i+1}) \frac{d E_{i, i+1}}{d \tau}=\frac{T}{T_g}\tau^{-1-\frac{T}{T_g}}
\end{equation}
for the PDF of $\tau$ and we identify $\alpha = T / T_g$. When $\alpha<1$ we get scale-free sojourn times.

The model is most easily analyzed, at least qualitatively, by considering the quenched transition rates. If the sojourn time to cross a barrier is $\tau$ then 
$W_{n,m} = \tau^{-1}$ is the transition rate from site $n$ to $m$ and are symmetrical $W_{n,m} = W_{m,n}$. 
It follows that the average waiting time at a site $n$ obeys $\tau_n^{-1} = (W_{n,n+1} + W_{n,n-1})/2$, which shows that any two neighboring sites are correlated.
The simulation is therefore constructed as follows: (i) at each time step, determine if we succeed in escape with probability $(W_{n,n+1} + W_{n,n-1})/2$. (ii) if we succeed, escape from right with probability $W_{n,n+1} / (W_{n,n+1} + W_{n,n-1})$, otherwise to the left. (iii) advance time by $\Delta t$, go back to step (i).
(*) Alternatively, we can say that the waiting time is taken from an exponential distribution with average $\langle \tau \rangle = 2 / (W_{n,n+1} + W_{n,n-1})$.

For the simulations of the CTRW model, the spatial and temporal disorder are independent and can therefore be treated separately. We first let the walker perform $N$ steps on the lattice and subsequently generate $N$ independent waiting times drawn from the L\'{e}vy-stable density $l_{\alpha,A,1}(\eta)$. The physical time is then obtained by summing these waiting times.
If the accumulated time remains smaller than the prescribed measurement time $t$, the walker is allowed to perform additional steps, and new independent waiting times are generated accordingly. This procedure is repeated until the total elapsed time reaches $t$.

In QTM, the landscape is randomized only once and subsequently remains fixed throughout the entire trajectory. This quenched disorder generates temporal correlations: upon revisiting a site, the particle experiences the same waiting time as during its first visit.
To implement this numerically, we proceed as follows. First, the walker performs $N$ steps in the underlying lattice. We then identify the set of distinct sites visited during these steps and assign waiting times only to those sites. The total physical time elapsed up to step $N$ is obtained by summing the waiting times associated with the visited sites, accounting for repeated visits through their previously assigned values. 
If the accumulated time is smaller than the prescribed measurement time $t$, the walker is allowed to perform additional steps. Waiting times are generated only for newly visited sites, while previously visited sites retain their originally assigned values. This procedure is repeated until the total elapsed time reaches the fixed measurement time $t$.

In the Comb model, the quenched teeth lengths $L$ are drawn from a heavy-tailed distribution $p(L)\sim L^{-1-\gamma}$ for $L\gg 1$ with $0<\gamma<1$. 
The walker starts at position $(x,y) = (0,0)$ where $x$ is the position on the back bone and $y$ is the position on the tooth. we follow the simulation procedure as in Ref.~\cite{havlin1987anomalous}.
When the walker is at the backbone ($y=0$), it moves along the $x$ direction with total probability $1/2$ (choosing right or left with equal probabilities), or enters the attached tooth with probability $1/2$. Importantly, the walker may reenter a tooth he just exited.
Once inside a tooth ($y>0$), it may go up or down with equal probabilities. At the reflecting tip of a tooth ($y=L$), the boundary condition implies that the walker either remains at $L$ with probability $1/2$ or steps to $L-1$ with probability $1/2$.
Time in the simulation is discrete and is identified with the number of steps taken by the walker.
For the simulation results showing the collapse onto the Mittag–Leffler distribution in the main text [Fig.~3], we used a lag time $\Delta$ large enough to include the effects of reaching the ends of the quenched teeth.

\end{document}